\documentclass[11pt,a4paper]{article}
\usepackage[T1]{fontenc}
\usepackage[utf8]{inputenc}
\usepackage{lmodern}
\usepackage[margin=24mm]{geometry}
\usepackage{amsmath,amssymb}
\usepackage{graphicx}
\usepackage{booktabs,longtable,array}
\usepackage{xcolor}
\usepackage{listings}
\usepackage{microtype}
\usepackage{url}

\usepackage[hidelinks]{hyperref}
\hypersetup{pdftitle={From Human Intent to Strategy Decisions: A Review Framework Integrating Large Language Models and Digital Twins},pdfauthor={Teng-Hsien Ko and Chin-Te LIN},pdfkeywords={Human intent; Strategy review; Large language models; Digital twin; Human-AI collaboration; Traceable verification}}
\title{From Human Intent to Strategy Decisions: A Review Framework Integrating Large Language Models and Digital Twins}
\author{Teng-Hsien Ko \and Chin-Te LIN\thanks{Chin-Te LIN, Associate Professor\newline ORCID: \href{https://orcid.org/0000-0002-4401-4792}{0000-0002-4401-4792}\newline Department of Mechanical Engineering, National Central University\newline Email: \href{mailto:chintelin@ncu.edu.tw}{\nolinkurl{chintelin@ncu.edu.tw}}\newline No. 300, Zhongda Rd., Zhongli Dist., Taoyuan City 320317, Taiwan (R.O.C.)}}
\date{\small Department of Mechanical Engineering, National Central University, Taiwan}
\begin{document}
\maketitle
\begin{abstract}
Automation systems must adapt to changing tasks, equipment states, and staffing conditions while providing evidence for human review. This study presents an intent-review framework integrating large language models, digital twins, and human decision-making, implemented for multi-line task adjustment. A Propose-Verify-Decide workflow translates operator intent into structured requirements, generates a bounded set of candidate strategies, and checks semantics, simulation execution, and operational constraints. Linked records preserve traceability from requests to verification evidence and decisions. Thirty fixed test records were evaluated using four virtual surgical-instrument sorting lines: 28 assessed the workflow and two assessed model generation. Eighteen workflow cases met expectations (64.29\%); three of ten strategy-workflow requests completed autonomously (30\%); and seven of eight invalid inputs were correctly intercepted (87.5\%). Analysis by anomaly source showed that the framework rejected two candidate batches that failed throughput constraints and paused recommendations in two cases with incomplete evidence. A mistranslated priority rule nevertheless reached simulation. These findings demonstrate handling of specific invalid inputs, throughput violations, and evidence gaps, and identify consistency between original intent and execution specifications as a key improvement area. The effect on overall erroneous decisions remains to be evaluated through matched comparisons, together with the influence of digital-twin fidelity using independent physical references.
\end{abstract}
\medskip

\section{Introduction}

Manufacturing automation assigns repetitive operations to equipment, but changes in product mix, urgent requests, equipment states, and personnel availability can alter the conditions under which a production plan was established. When a line must prioritize an urgent task or suspend some operations following an anomaly, the system must determine which tasks can continue, which constraints must remain unchanged, and how the adjustment affects other lines. Research on reconfigurable manufacturing systems (RMSs) treats changes in capacity, functionality, and system structure as a foundation for manufacturing flexibility \cite{r1,r2}. Software-defined manufacturing extends this capability by managing equipment behavior and process configuration through software \cite{r3,r4,r5}. Software-defined value-stream process systems connect product requirements, machine capabilities, and digital planning, identifying automated reconfiguration and shortages of skilled personnel as research motivations \cite{r6}. These architectures provide mechanisms for adapting equipment and processes, but actual changes still require people to interpret the current situation, translate requirements into settings, check constraints, and assess consequences.

Large language models (LLMs) provide a natural-language interface for translating such requirements. Studies of manufacturing task planning and multi-agent manufacturing have explored converting human instructions into task steps, process descriptions, and coordination messages \cite{r7,r8,r9,r10}. Operators can thereby express the purpose of a change while the system organizes the request into data that downstream modules can process. Relevant information may include process rules, equipment limits, exception-handling procedures, and historical records. However, coherent text or structurally valid output does not establish that a generated strategy is appropriate for the current production setting. Even a semantically correct scenario may be infeasible because of equipment capabilities, instrument placement, or throughput constraints. Skill-feasibility and formal-verification studies have addressed the connection between language planning and executable capabilities, as well as the conformity of generated programs to task specifications \cite{r11,r12,r13}. For the multi-line task adjustments considered here, requirement semantics, scenario-execution results, and operational constraints must be brought together in a review workflow that supports judgments about strategy applicability.

The central research question is how operator-requested task changes can be organized into a strategy-review workflow that is inspectable, allows rejection, and preserves traceability, using a local model and an existing simulation environment. The study focuses on providing practical adjustment tools and decision evidence when a limited number of engineering or operational personnel must repeatedly adapt an automated system.

The proposed system combines an LLM and a digital twin through three stages: Propose, Verify, and Decide. The proposal stage organizes natural-language intent and production states into a bounded set of candidate strategies. The verification stage converts each candidate into a scenario specification, executes it in the digital twin, and checks mandatory constraints. The decision stage presents the recommended strategy, its provenance, performance indicators, and supporting evidence to the operator. A digital twin allows strategies to be evaluated through an observable process before they affect physical equipment \cite{r14,r15,r16,r17}. In this division of responsibilities, the LLM assists with information processing and strategy formulation, the digital twin provides execution evidence for a specified scenario, and people retain contextual judgment and responsibility for adoption.

The contributions are threefold. First, the system links operator intent, task-requirement versions, state records, candidate strategies, scenario specifications, and execution evidence, enabling strategy provenance and verification outcomes to be traced. Second, it implements the Propose-Verify-Decide workflow by filtering a bounded candidate set against mandatory constraints, ranking eligible candidates by throughput-target attainment, and retaining revision and cancellation branches. Third, a fixed test suite covering ordinary requests, invalid inputs, and model generation distinguishes semantic correctness, structural validity, simulation execution, constraint satisfaction, and outcome interpretation, while measuring the time required at different stages. These contributions concern system integration and experimental analysis.

\section{Related Work}

LLMs can translate natural language into task representations and interact with external systems through tool interfaces, motivating their use in manufacturing agents and robot planning. Lim et al. studied LLM-enabled multi-agent manufacturing systems \cite{r7}, Tsushima et al. investigated task planning for factory robots \cite{r8}, and Fan et al. examined embodied intelligence for industrial robots \cite{r9}. Ni et al.'s LLMAPM converts user descriptions into structured manufacturing task flows \cite{r10}. These studies offer different integration approaches for translating language into operational representations. External knowledge and tools also play important roles. Retrieval-augmented generation incorporates retrievable knowledge into generation \cite{r18}; ReAct links reasoning, actions, and external observations \cite{r19}; and Toolformer studies how models can use tools \cite{r20}. SayCan connects high-level language proposals with robot-skill affordances \cite{r11}. Bayat et al. use code-generation and checking agents to connect natural-language requirements to symbolic-control tools \cite{r12}. Yang et al. convert generated robot programs into automata, verify them against safety specifications, and use the results to support model refinement \cite{r13}. Together, these approaches cover coordination, planning, and process generation while demonstrating the need for external capability or specification checks.

A digital twin (DT) represents equipment and operational states through a digital model, allowing candidate strategies to be evaluated without directly changing a physical production line. Qamsane et al. proposed a methodology for developing and implementing manufacturing DT solutions \cite{r16}, while Leng et al. reviewed their role in smart-manufacturing system design \cite{r21}. For production-line adjustment, Yang et al. used DTs to assess flexible and reconfigurable automotive-part lines \cite{r14}, Fu et al. studied dynamic line reconfiguration under disturbances \cite{r15}, and Huang et al. combined DTs with deep reinforcement learning for reconfiguration planning \cite{r22}. These studies show that digital models can evaluate candidate configurations and strategies in addition to displaying equipment states. In task execution and human-machine interaction, Li et al. applied a DT to robot task replanning and human-robot control \cite{r17}, and Leng et al. integrated monitoring and simulation for reconfigurable manufacturing systems \cite{r23}. DTs can thus provide event observations and quantitative evidence. Such evidence must nevertheless be interpreted in relation to model fidelity. Zhang et al. propose a multilevel DT modeling method with fidelity evaluation \cite{r24}. The applicability of DT evidence to physical operations therefore requires further confirmation.

Several studies integrate LLMs with DTs. FactoryFlow examines resilience and human oversight in LLM-assisted DT modeling, separating structural modeling from parameter fitting and controlling modeling errors through constrained intermediate representations and prevalidated components \cite{r25}. The present study similarly uses an intermediate representation, but translates task changes into executable scenarios in an existing environment. MAKA separates intent routing, tool-based numerical analysis, knowledge graphs, and verification agents to check physical plausibility, safety boundaries, and provenance before human review. Its case study concerns evidence integration and compensation recommendations for aerospace-blade machining \cite{r26}. It shares the present study's emphasis on traceable verification and human decision-making; here, the focus is on task changes across multiple robot-operated lines and comparable evidence from candidate-strategy simulations. Cruz et al. integrated LLMs and AutoML in the GAMHE 5.0 pilot line, using LLM-generated code to prepare datasets and AutoML to construct predictive models for process optimization \cite{r27}. Their work emphasizes the connection between data, models, and optimization, whereas this study focuses on preserving operator intent and decision history during task adjustment. Table~\ref{tab:1} compares the main objects and verification emphases of these approaches.

\begin{table}[htbp]\centering
\begingroup
\small\setlength{\tabcolsep}{4pt}\renewcommand{\arraystretch}{1.13}
\caption{Scope and verification focus of related approaches.}\label{tab:1}\medskip
\begin{tabular}{@{}>{\raggedright\arraybackslash}p{0.16000\dimexpr\linewidth-6\tabcolsep\relax}>{\raggedright\arraybackslash}p{0.23000\dimexpr\linewidth-6\tabcolsep\relax}>{\raggedright\arraybackslash}p{0.27000\dimexpr\linewidth-6\tabcolsep\relax}>{\raggedright\arraybackslash}p{0.34000\dimexpr\linewidth-6\tabcolsep\relax}@{}}
\toprule
\textbf{Study} & \textbf{Main object} & \textbf{Verification or evaluation focus} & \textbf{Relationship to this study} \\
\midrule
LLMAPM \cite{r10} & Natural language to manufacturing processes & Process generation and platform integration & This study also addresses scenario evidence and candidate review. \\ \addlinespace[3pt]
SayCan \cite{r11} & High-level instructions to robot skills & Skill affordances and executability & This study considers multiple lines, constraint checks, and strategy comparison. \\ \addlinespace[3pt]
Symbolic control and program verification \cite{r12,r13} & Natural language to control specifications or programs & Specification conformance and formal checks & This study evaluates candidates through scenario execution and constraint checks. \\ \addlinespace[3pt]
FactoryFlow \cite{r25} & Natural language to twin models & Intermediate representations, modeling errors, and oversight & This study uses an existing scene to verify task strategies. \\ \addlinespace[3pt]
MAKA \cite{r26} & Machining evidence to compensation recommendations & Tool analysis, physical plausibility, and provenance & Both emphasize evidence and human decisions, with different applications and evaluation units. \\ \addlinespace[3pt]
GAMHE 5.0 \cite{r27} & Data, predictive models, and optimization & Code functionality and model evaluation & This study focuses on semantics, execution, and review during task changes. \\ \addlinespace[3pt]
This study & Operator intent to multi-line candidate strategies & Conditional checkpoints, placement and throughput evidence, and outcome interpretation & Local prototype, bounded candidates, and fixed-case analysis. \\ \addlinespace[3pt]
\bottomrule\end{tabular}\endgroup
\end{table}

Human-robot task-allocation research considers the assignment of personnel, machines, and tasks. Fusaro et al. integrated role allocation and task planning for mixed human-robot teams \cite{r28}, Lippi et al. proposed a task-allocation framework for human-multi-robot collaboration \cite{r29}, and Lamon et al. assigned industrial-assembly roles according to capabilities \cite{r30}. Frering et al. combined belief-desire-intention agents with LLMs to study reliable human-robot interaction and explainability \cite{r31}. These studies indicate that collaboration quality depends both on who can perform a task and on how people understand intelligent-system behavior.

The reviewed studies provide complementary methods for natural-language conversion, candidate evaluation, evidence verification, and human review. The integration problem addressed here is how to preserve the original intent across requirement conversion, strategy generation, simulation execution, and result review after an operator requests a multi-line change, while tracing rejection or revision to a specific processing stage.

Accordingly, this study connects a local model, an existing DT scene, and human review through a traceable data chain. Mandatory constraints are checked within a bounded candidate set before a recommendation is issued. The scope is workflow integration and verification for task adjustment; automatic twin construction, globally optimal scheduling, and formal safety guarantees are outside this scope.

\section{Method}

\subsection{Governance Concept}

\begin{figure}[htbp]\centering
\includegraphics[width=.78\linewidth,height=.77\textheight,keepaspectratio]{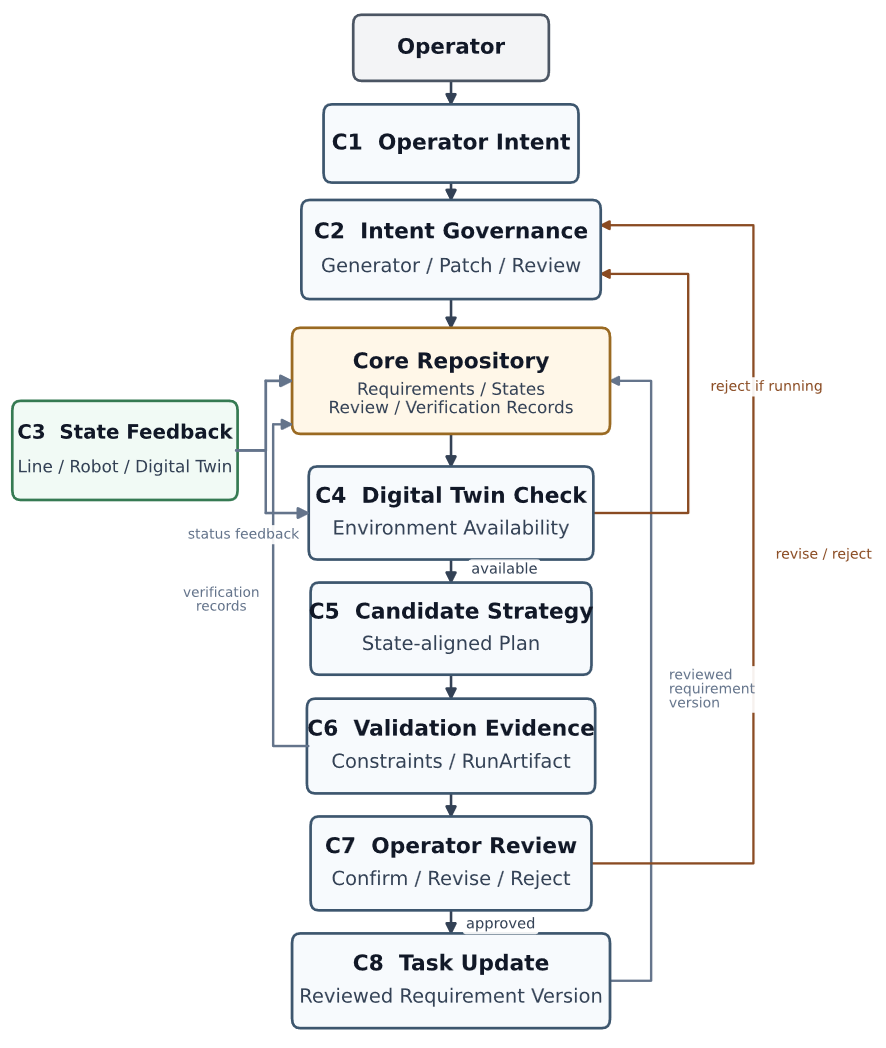}
\caption{Closed-loop governance concept.}\label{fig:1}
\end{figure}

\begin{figure}[htbp]\centering
\includegraphics[width=\linewidth,height=.78\textheight,keepaspectratio]{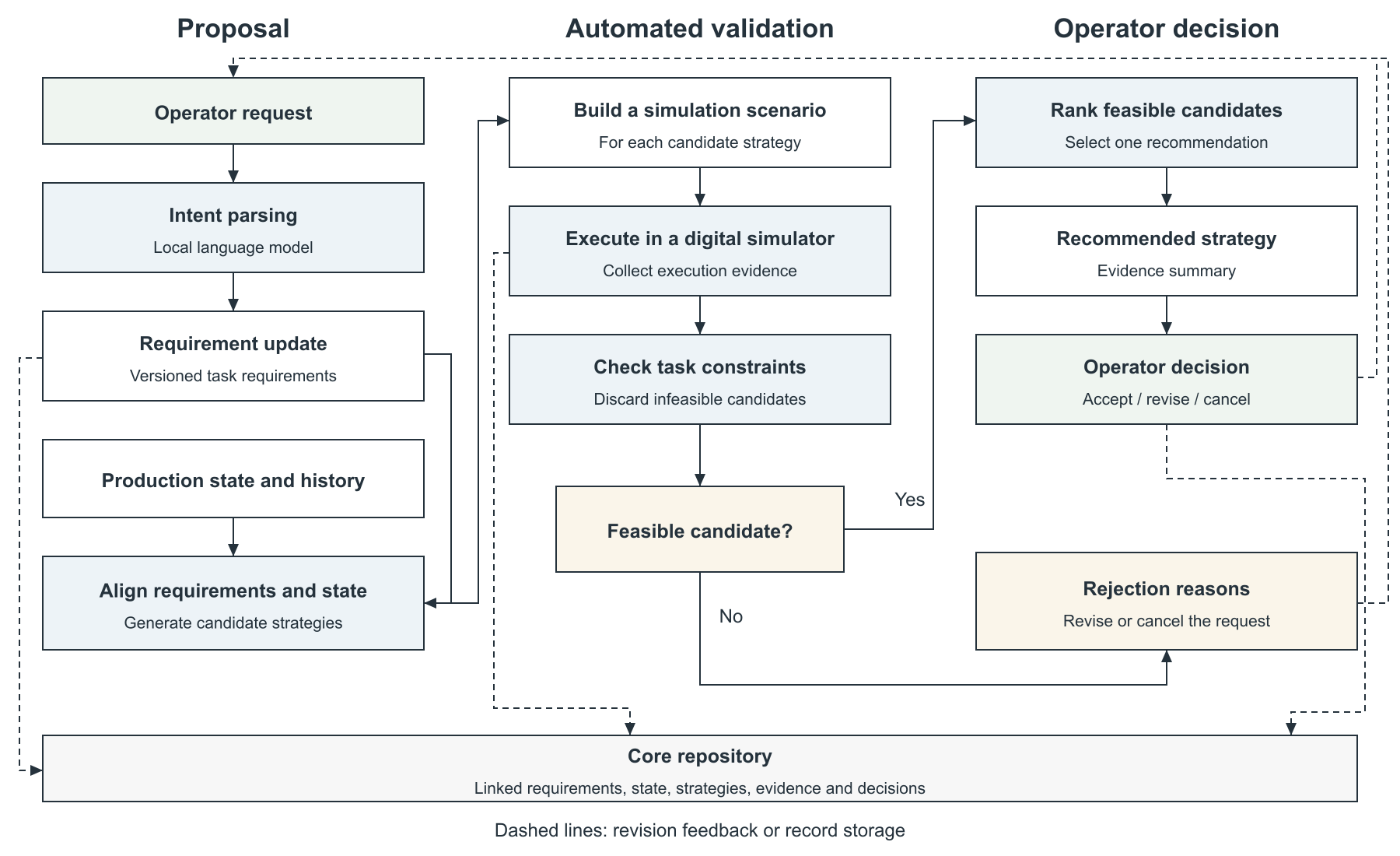}
\caption{Workflow from requirement changes and candidate verification to operator decisions. After filtering and ranking, the system presents a recommendation and evidence summary. If no candidate qualifies, it reports the reasons and requests revision or cancellation.}\label{fig:2}
\end{figure}

The method organizes task adjustment into proposal, verification, and decision-making. The proposal stage parses operator intent, establishes a requirement version, and aligns it with production states. The verification stage executes candidate strategies in the DT, checks mandatory constraints, and compiles evidence. The decision stage allows the operator to accept, revise, or cancel a proposal based on the recommended strategy and its evidence. Structured requirements, candidate configurations, and check results are retained between the original request and strategy adoption so that decisions remain traceable. Figure~\ref{fig:1} illustrates the closed-loop governance concept. C3-C6 form the execution and verification feedback path, whereas C1, C2, C7, and C8 form the operator-intent and review path. The method separates three levels of judgment: the LLM interprets natural language and produces structured requirements; the DT executes scenarios and generates scene, placement, and throughput evidence; and the operator makes an adoption decision. Linking these levels creates traceable stages and prevents a single module's judgment from becoming the sole basis for deployment. Figure~\ref{fig:2} shows the workflow.

\subsection{System Architecture and Data Flow}

\begin{figure}[htbp]\centering
\includegraphics[width=\linewidth,height=.78\textheight,keepaspectratio]{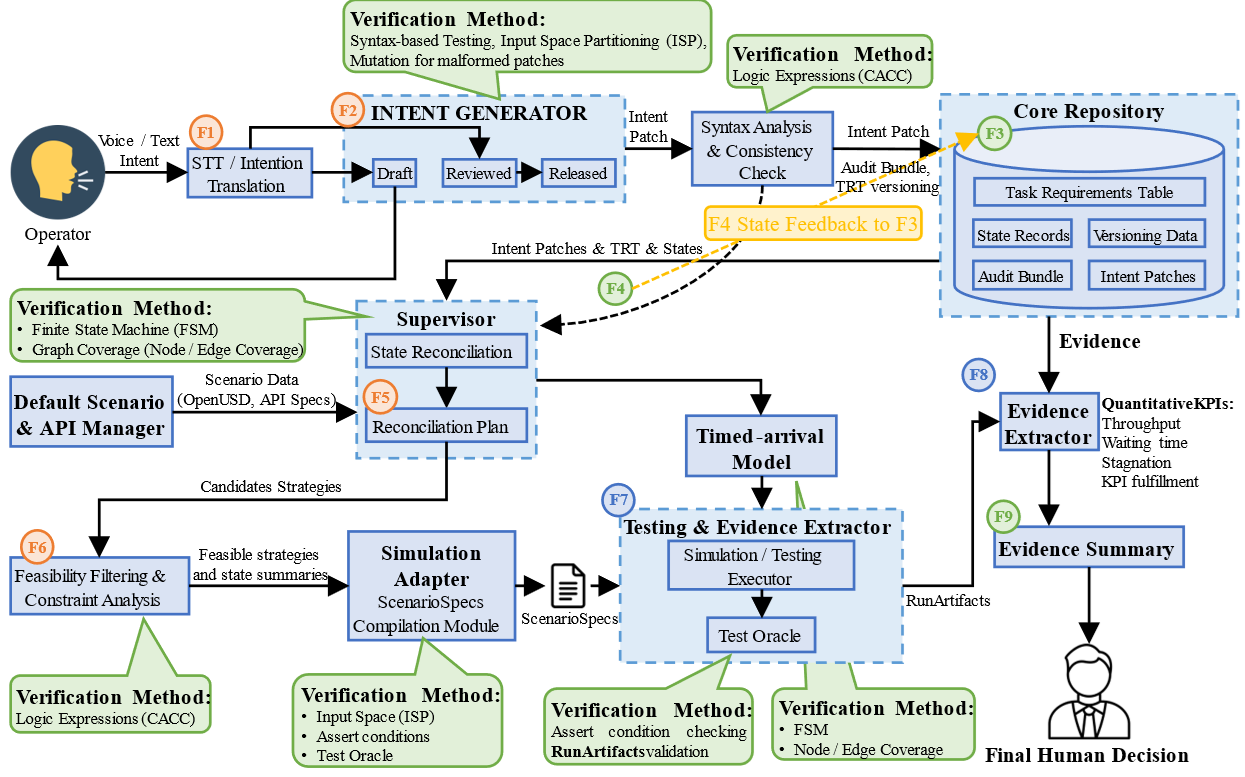}
\caption{System modules and data relationships. Solid lines indicate requests, configurations, or results exchanged between modules; dashed lines indicate reads and writes to the Core Repository.}\label{fig:3}
\end{figure}

Figure~\ref{fig:3} shows the system modules and data flow. The system receives two types of input: operator intent concerning task adjustments and structured state data obtained through production-line or DT interfaces. Intent enters the intent-governance workflow. State data include task status, instrument positions, anomalies, intervention records, and verification results and are retained through a state-recording workflow. Both types are linked in the Core Repository to support candidate generation. F1-F9 identify the main data-flow nodes in Figure~\ref{fig:3}.

Operator intent enters at F1. At F2, the Intent Generator uses the local LLM to identify the request type and required information. Supported types include task changes, configuration or state queries, requirement approval, strategy-adoption decisions, cancellation, and general dialogue. Task changes require an operator identifier, a reason, a target scope, and the requested modification. Missing or invalid fields should trigger clarification or rejection. A validated Intent Patch is linked to the task-requirement version in F3 through the requirement-confirmation workflow.

The Supervisor reads task requirements and state records from the Core Repository and reconciles them. Production-line or DT states are written back through F4, providing traceable state information for candidate generation. After reconciliation, candidate strategies are produced at F5.

At F6, the Feasibility and Constraint Analysis module performs preliminary checks against the scene and rules. The Simulation Adapter converts eligible candidates into scenario specifications for simulation execution and test evaluation in the Testing \& Evidence Extractor at F7. The digital simulator produces execution status, event records, instrument-placement data, and verification reports recording passed checks, warnings, and failures. At F8, the Evidence Extractor organizes the results into execution-evidence records (RunArtifacts) for candidate screening. The resulting Evidence Summary at F9 supports operator review.

The architecture comprises intent-governance, state-feedback, and verification-evidence paths. Identifiers link requirement versions, candidate strategies, scenario specifications, and execution evidence, enabling the origin and verification results of each strategy to be traced.

\subsection{Candidate Generation, Verification, and Decision-Making}

Each strategy batch contains multiple candidates that vary permitted strategy fields while preserving the operator-specified line scope, target instruments, and throughput constraints. This bounded set supports comparison of alternatives; the selected candidate meets the constraints and ranks highest within that batch. Algorithm 1 summarizes generation, verification, and selection. The Evidence Extractor organizes execution evidence, and the StrategySelector filters candidates according to evidence completeness, operational outcomes, and mandatory constraints. Eligible candidates are then ranked by the extent to which line throughput meets its target. The highest-ranked candidate is presented with its change scope, constraint-check results, and applicability conditions. If no candidate qualifies, the system reports the reasons and requests revision or cancellation. Candidate evaluations and operator decisions are stored in the repository.

\begin{figure}[p]
\textbf{Algorithm 1. Bounded strategy generation, verification, and selection.}\par\medskip
\begin{lstlisting}
Input: operator request, current requirement version, state records,
       scene contract, validation rules, throughput constraints
Output: selected strategy or refinement request, evidence summary

1:  status <- DigitalTwinEnvironment.getSimulationStatus()
2:  if Supervisor.requiresCheckpoint(status) then
3:      return CoreRepository.saveBusyRejection(status)
4:  end if
5:  decision <- IntentService.classifyAndNormalize(operator request)
6:  if decision.requiresClarification() then
7:      return IntentService.requestMissingSlots(decision)
8:  end if
9:  validation <- PatchValidationService.validate(decision.normalizedRequest)
10: if validation.failed() then
11:     return CoreRepository.saveVerificationRecord(validation)
12: end if
13: release <- ReleaseManager.prepareRelease(validation.patch)
14: state <- CoreRepository.loadStateRecords(release.scope)
15: plan <- Supervisor.reconcile(release.requirements, state)
16: strategyBatch <- CandidateStrategyGenerator.generateBatch(plan, count = 3)
17: for each strategy in strategyBatch do
18:     scenario <- ScenarioAdapter.generate(strategy, scene contract)
19:     if scenario.validationFailed() then
20:         CoreRepository.saveVerificationRecord(scenario.result)
21:         continue
22:     end if
23:     run <- DigitalTwinEnvironment.run(scenario)
24:     evidence <- EvidenceExtractor.summarize(run, scenario)
25:     CoreRepository.saveEvidence(evidence)
26: end for
27: eligible <- StrategySelector.applyHardGates(strategyBatch.evidence, throughput constraints)
28: if eligible.isEmpty() then
29:     return StrategySelector.createRefinementRequest(strategyBatch.evidence)
30: end if
31: ranked <- StrategySelector.rankByThroughputAttainment(eligible)
32: summary <- DecisionSummaryGenerator.create(ranked, strategyBatch.evidence)
33: decision <- Operator.review(summary)
34: CoreRepository.saveReviewRecord(decision, summary)
35: return ranked.first(), summary
\end{lstlisting}\end{figure}

\subsection{CSSD Digital-Twin Verification Scenario}

The verification scene represents instrument sorting and set assembly in a central sterile supply department (CSSD). Studies of sterile processing describe coordination among work steps, personnel, and information \cite{r32,r33,r34}, while workload and productivity assessment require consideration of work activities and time requirements \cite{r35,r36}. Automated operations also require human intervention when exceptions arise. In this setting, a strategy's value depends not only on throughput but also on whether people can inspect the change scope, failure reasons, and waiting costs. A CSSD DT was therefore constructed as the virtual verification environment.

\subsection{Evidence Interpretation and Human Roles}

The decision summary first reports evidence completeness and satisfaction of mandatory constraints, followed by instrument completion, placement results, anomaly events, and time costs. A strategy that violates a mandatory constraint must not be presented as eligible for adoption. Even an eligible strategy requires an operator decision based on task objectives and local conditions.

Three human activities are distinguished: the operator's adoption decision, final human review of complete evidence at CP6, and retrospective case review by the researchers. The first concerns responsibility for adoption; CP6 checks the output and evidence of cases reaching that stage; retrospective review determines whether each case meets its expected behavior by comparing the original request, expected outcome, and stopping point. Correctly rejecting an infeasible strategy may satisfy a test objective, whereas stopping because of an error or an external cancellation does not by itself constitute correct rejection.

\section{Experimental Design}

\subsection{Implementation Environment}

Table~\ref{tab:2} lists the hardware and software configuration. n8n provides the operator interface and workflow orchestration, Python connects the modules, SQLite stores execution records, and JSON files retain scenario specifications and RunArtifacts. A local LLM is used as the inference backend so that requirements, states, and execution data can be processed within a controlled service environment. The evaluation assesses workflow usability under the specified hardware and model configuration.

\begin{table}[htbp]\centering
\begingroup
\small\setlength{\tabcolsep}{4pt}\renewcommand{\arraystretch}{1.13}
\caption{Main implementation environment.}\label{tab:2}\medskip
\begin{tabular}{@{}>{\raggedright\arraybackslash}p{0.24000\dimexpr\linewidth-2\tabcolsep\relax}>{\raggedright\arraybackslash}p{0.76000\dimexpr\linewidth-2\tabcolsep\relax}@{}}
\toprule
\textbf{Component} & \textbf{Configuration and purpose} \\
\midrule
Inference server & NVIDIA RTX A6000 48 GB; Intel Xeon E5-2643 v4; 32 GB RAM; Ubuntu 20.04 LTS \\ \addlinespace[3pt]
Local experiment host & NVIDIA GeForce RTX 3060; Intel Core i5-12400; 32 GB RAM; Windows 10 \\ \addlinespace[3pt]
Primary model & \path{cyankiwi/gemma-4-26B-A4B-it-AWQ-8bit} \\ \addlinespace[3pt]
Inference service & vLLM 0.21.0 \\ \addlinespace[3pt]
Workflow and integration & n8n 2.21.7; task-requirement management service; Python 3.11 \\ \addlinespace[3pt]
Physics simulation & Isaac Sim 5.1.0 \\ \addlinespace[3pt]
Evidence storage & SQLite execution records and JSON RunArtifacts \\ \addlinespace[3pt]
\bottomrule\end{tabular}\endgroup
\end{table}

\begin{figure}[htbp]\centering
\includegraphics[width=\linewidth,height=.78\textheight,keepaspectratio]{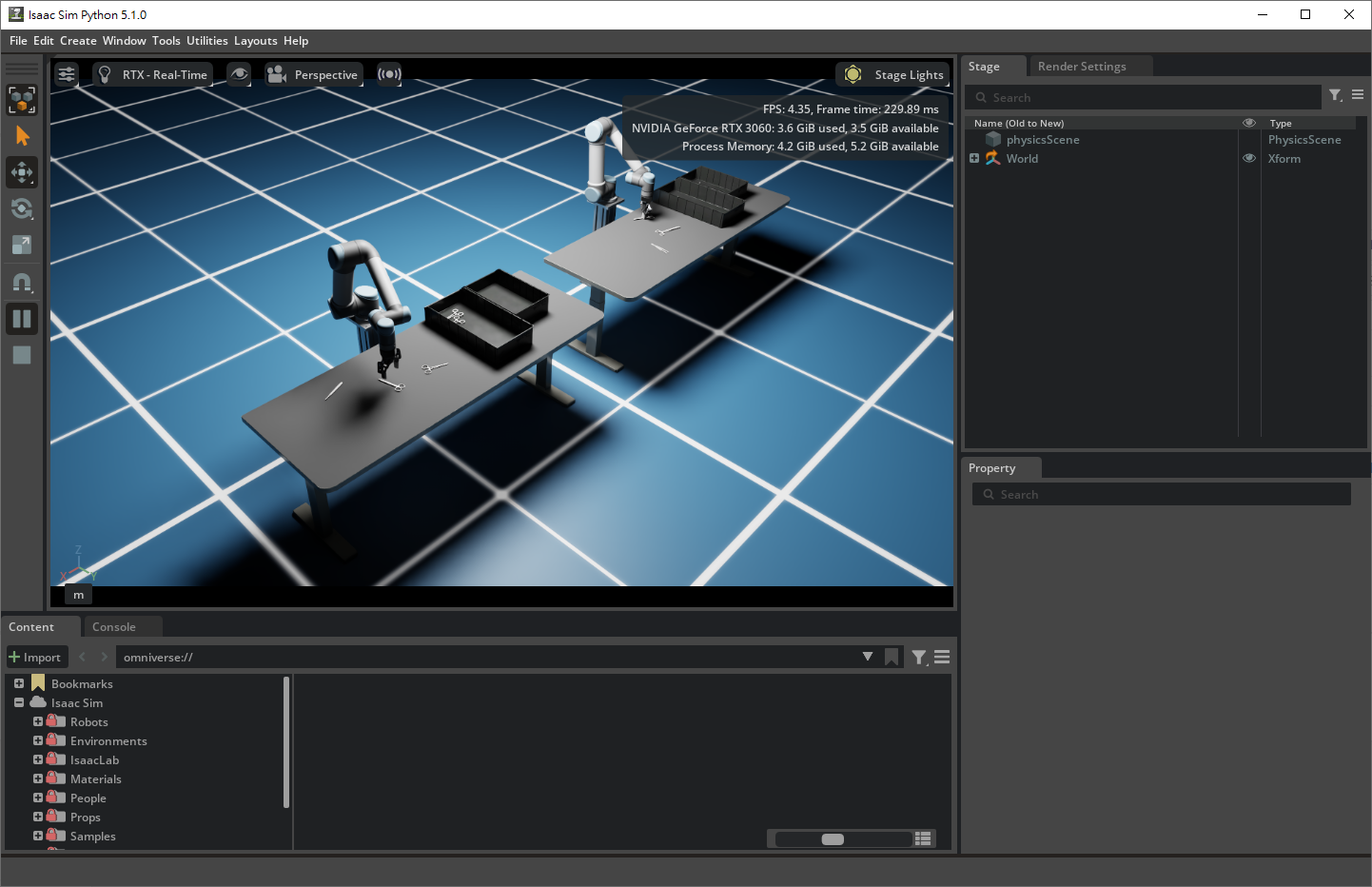}
\caption{Instrument-sorting workstations in Isaac Sim, showing the simulated equipment and workspace.}\label{fig:4}
\end{figure}

The verification environment represents one operator and four automated CSSD lines in Isaac Sim 5.1.0. Figure~\ref{fig:4} shows the instrument-sorting scene. Each line has a UR5 robot workstation that picks specified categories of surgical instruments and places them at designated locations. Here, set assembly means sorting and placing instruments according to a predefined list. The evaluation primarily concerns text requests and parameterized intervention workflows; real operators' speech-recognition or physical-intervention performance is not evaluated.

\subsection{Test Design}

\subsubsection{Checkpoints}

\begin{figure}[htbp]\centering
\includegraphics[width=\linewidth,height=.78\textheight,keepaspectratio]{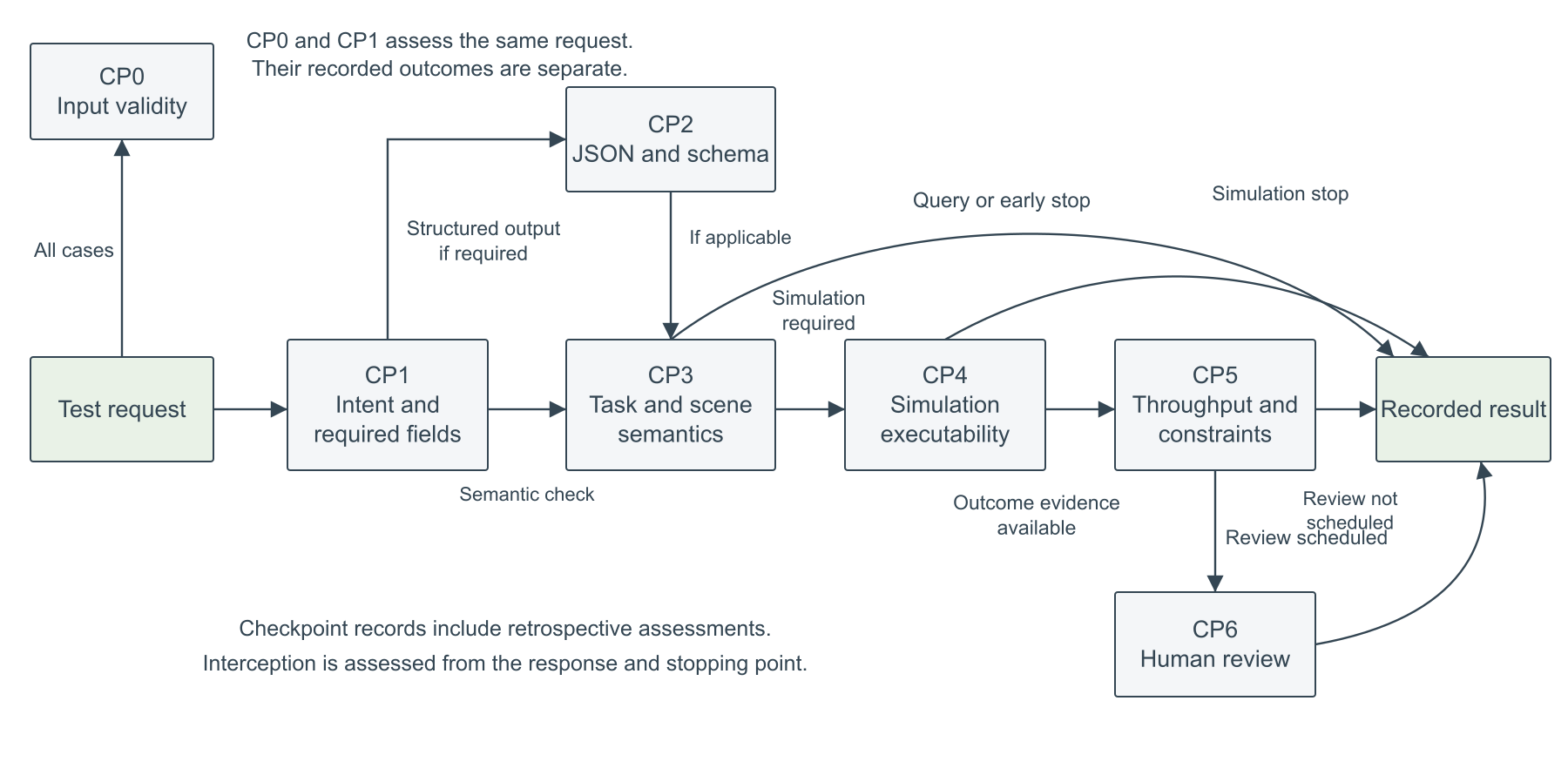}
\caption{Checkpoint routes by request type. CP0 and CP1 may assess the same request independently. CP6 records final human review for cases with complete evidence. All cases are also subject to retrospective outcome interpretation.}\label{fig:5}
\end{figure}

Observable checkpoints (CPs) are defined for requirement conversion, simulation verification, and result review. Figure~\ref{fig:5} shows CP0-CP6 and their applicable routes by request type; Table~\ref{tab:3} defines their criteria. CP0 evaluates the original input, while CP1 evaluates intent and required fields. Invalid, incomplete, or contradictory requests should trigger rejection or clarification. Cases requiring structured output proceed to CP2 for schema checks, and CP3 verifies that line scope, instruments, priority rules, and query scope preserve the intended meaning. Queries may end after semantic checking. Requests requiring strategy verification proceed to CP4 for simulation execution and evidence completeness, followed by CP5 for throughput, placement, and other constraints. Only cases with complete evidence that are scheduled for human review reach CP6. The evaluated sets therefore differ across checkpoints rather than forming a single sequential elimination funnel. Checkpoint records also include evaluations made from preserved records: a recorded failure does not necessarily mean execution was blocked at that time. Correct interception must be assessed against the actual response and stopping point.

\begin{table}[htbp]\centering
\begingroup
\small\setlength{\tabcolsep}{4pt}\renewcommand{\arraystretch}{1.13}
\caption{Checkpoint purposes, applicability conditions, and pass criteria.}\label{tab:3}\medskip
\begin{tabular}{@{}>{\raggedright\arraybackslash}p{0.06500\dimexpr\linewidth-6\tabcolsep\relax}>{\raggedright\arraybackslash}p{0.25500\dimexpr\linewidth-6\tabcolsep\relax}>{\raggedright\arraybackslash}p{0.38000\dimexpr\linewidth-6\tabcolsep\relax}>{\raggedright\arraybackslash}p{0.30000\dimexpr\linewidth-6\tabcolsep\relax}@{}}
\toprule
\textbf{CP} & \textbf{Entry condition} & \textbf{Pass criterion} & \textbf{Subsequent route} \\
\midrule
CP0 & An operator request is received. & Input is interpretable and belongs to a supported query or change-request route. & Record input validity; CP1 may independently assess the same request. \\ \addlinespace[3pt]
CP1 & Intent classification or required-field checking is needed. & Intent and required fields conform, or missing information triggers correct clarification or rejection. & Proceed to CP2 or CP3 according to output type. \\ \addlinespace[3pt]
CP2 & JSON or another structured representation is expected. & Output meets the specified schema, including required fields and types. & Proceed to CP3 when semantic checking is required. \\ \addlinespace[3pt]
CP3 & The case concerns tasks, lines, instruments, priorities, or query scope. & Task and scenario semantics match the case specification. & Proceed to CP4 if simulation is needed; queries may end here. \\ \addlinespace[3pt]
CP4 & An executable scenario specification has been produced and the DT started. & The scenario executes successfully and produces the required simulation evidence. & Proceed to CP5 when throughput evidence is available. \\ \addlinespace[3pt]
CP5 & throughput, placement, priority, or safety-condition evidence is available. & All applicable constraints meet the case specification. & Proceed to CP6 if human review is scheduled. \\ \addlinespace[3pt]
CP6 & Complete evidence is available for human result review. & A researcher checks the records to confirm that the output, stopping point, and rejection behavior meet the specification. & Store the review result; this is not a physical-deployment decision. \\ \addlinespace[3pt]
\bottomrule\end{tabular}\endgroup
\end{table}

\subsubsection{Test Cases}

The researchers acted as operators during testing, submitting requests, confirming requirements, and responding to the system. They subsequently reviewed the records to determine whether each case met its expected behavior. Table~\ref{tab:4} summarizes the design and outcomes. A test case is one record with an input and expected behavior; a test group collects cases sharing an evaluation purpose. TC1-TC5 assess the system workflow. TC6 assesses repeated generation by one model, and TC7 compares models. These two groups report structure, semantics, and completion status separately and are excluded from the case-conformance rate for TC1-TC5. Gemma, Qwen, and Llama are evaluated; their exact identifiers, settings, and results appear in Table~\ref{tab:5}. The evaluation distinguishes the scope of a throughput requirement from the simulation scope. An output passes strict semantic checking only if it preserves the requested scopes and produces the expected specification ready for review.

\begin{table}[htbp]\centering
\begingroup
\footnotesize\setlength{\tabcolsep}{4pt}\renewcommand{\arraystretch}{1.13}
\caption{Test groups and retrospective case-review outcomes. Model-generation outcomes are reported separately. Here, n is the number of records; Met and Not met refer to case expectations.}\label{tab:4}\medskip
\begin{tabular}{@{}>{\raggedright\arraybackslash}p{0.08000\dimexpr\linewidth-12\tabcolsep\relax}>{\raggedright\arraybackslash}p{0.10000\dimexpr\linewidth-12\tabcolsep\relax}>{\raggedright\arraybackslash}p{0.18000\dimexpr\linewidth-12\tabcolsep\relax}>{\raggedright\arraybackslash}p{0.30000\dimexpr\linewidth-12\tabcolsep\relax}>{\raggedright\arraybackslash}p{0.06500\dimexpr\linewidth-12\tabcolsep\relax}>{\raggedright\arraybackslash}p{0.17000\dimexpr\linewidth-12\tabcolsep\relax}>{\raggedright\arraybackslash}p{0.10500\dimexpr\linewidth-12\tabcolsep\relax}@{}}
\toprule
\textbf{Group} & \textbf{Cases} & \textbf{Main purpose} & \textbf{Procedure} & \textbf{n} & \textbf{Met} & \textbf{Not met} \\
\midrule
TC1 & T01-T08 & Intent to executable strategy & CP0-CP4; CP5 when evidence is produced; CP6 when review is reached & 8 & 3 & 5 \\ \addlinespace[3pt]
TC2 & T09-T17 & Configuration, state, and evidence queries & Compare replies with fixed query requirements & 9 & 5 & 4 \\ \addlinespace[3pt]
TC3 & T18-T19 & Simulation, throughput, and timing evidence & Execute virtual scenarios and collect quantitative evidence & 2 & 2 & 0 \\ \addlinespace[3pt]
TC4 & T20-T27 & Invalid-input interception & Inject missing, contradictory, or unsupported values & 8 & 7 & 1 \\ \addlinespace[3pt]
TC5 & T28 & Lifecycle and outcome records & Reconcile operation, verification, and review events & 1 & 1 & 0 \\ \addlinespace[3pt]
TC6 & T29 & Single-model variation & Repeat the same input three times with one model & 1 & Separate model evaluation & -- \\ \addlinespace[3pt]
TC7 & T30 & Cross-model generation & Repeat the same input three times per model & 1 & Separate model evaluation & -- \\ \addlinespace[3pt]
\bottomrule\end{tabular}\endgroup
\end{table}

\begin{table}[htbp]\centering
\begingroup
\footnotesize\setlength{\tabcolsep}{4pt}\renewcommand{\arraystretch}{1.13}
\caption{Model configurations and repeated-generation results for a single input.}\label{tab:5}\medskip
\begin{tabular}{@{}>{\raggedright\arraybackslash}p{0.34000\dimexpr\linewidth-10\tabcolsep\relax}>{\raggedright\arraybackslash}p{0.15000\dimexpr\linewidth-10\tabcolsep\relax}>{\raggedright\arraybackslash}p{0.15000\dimexpr\linewidth-10\tabcolsep\relax}>{\raggedright\arraybackslash}p{0.12000\dimexpr\linewidth-10\tabcolsep\relax}>{\raggedright\arraybackslash}p{0.12000\dimexpr\linewidth-10\tabcolsep\relax}>{\raggedright\arraybackslash}p{0.12000\dimexpr\linewidth-10\tabcolsep\relax}@{}}
\toprule
\textbf{Model} & \textbf{Quantization} & \textbf{Temperature / top-p} & \textbf{Structure valid} & \textbf{Strict semantic pass} & \textbf{Mean generation time} \\
\midrule
\path{cyankiwi/gemma-4-26B-A4B-it-AWQ-8bit} & AWQ 8-bit & 1.0 / 0.95 & 100\% (3/3) & 0\% (0/3) & 3.68 s \\ \addlinespace[3pt]
\path{Qwen/Qwen3.6-35B-A3B-FP8} & FP8 & 1.0 / 0.95 & 100\% (3/3) & 66.67\% (2/3) & 29.59 s \\ \addlinespace[3pt]
\path{meta-llama/Llama-3.1-8B-Instruct} & Unquantized & 0.6 / 0.9 & 100\% (3/3) & 0\% (0/3) & 8.01 s \\ \addlinespace[3pt]
\bottomrule\end{tabular}\endgroup
\end{table}

\subsection{Evaluation Metrics}

\subsubsection{Workflow Metrics}

Workflow evaluation includes checkpoint pass rates, case conformance, autonomous strategy-workflow success, candidate-batch selection, query success, and correct rejection, together with response and verification times. Table~\ref{tab:6} in Section 5.2 combines the evaluated populations, criteria, and results. For a proportion metric $q$ with a binary outcome,

\begin{equation}
R_q=\frac{n_q}{N_q}\times100\%.
\end{equation}

Here, $q$ denotes a proportion metric in Table~\ref{tab:6}, $N_q$ is the number of cases or batches included according to that metric's definition, $n_q$ is the number meeting its criterion, and $R_q$ is the percentage. For CP0-CP5, $N_q$ includes only cases evaluated at that checkpoint; for CP6, it comprises the four records that actually reached human review. For correct rejection, $n_q$ counts inputs rejected in accordance with their case specifications. A rate is computed only when $N_q>0$; otherwise, it is not applicable. Placement metrics use instrument records as the observation unit, as described in Section 4.3.2.

Time to the first reviewable response is defined as

\begin{equation}
T_{\mathrm{first}}=t_{\mathrm{first\ reply}}-t_{\mathrm{request}}.
\end{equation}

The request-receipt time is $t_{\mathrm{request}}$, and $t_{\mathrm{first\ reply}}$ is the time of the first response sufficient to determine the next action. Their difference, $T_{\mathrm{first}}$, is measured in seconds. This interval excludes operator reading or deliberation, and the first response is not necessarily the final strategy recommendation. Simulation-verification time is

\begin{equation}
T_{\mathrm{sim}}=(t_{\mathrm{evidence}}-t_{\mathrm{scenario}})-\Delta t_{\mathrm{startup}}.
\end{equation}

Here, $t_{\mathrm{scenario}}$ marks entry into the simulation-execution lifecycle, and $t_{\mathrm{evidence}}$ marks completion of execution evidence. Their difference is elapsed wall-clock time. The initial simulator-startup interval, $\Delta t_{\mathrm{startup}}$, is subtracted to obtain $T_{\mathrm{sim}}$. This interval measures simulation execution and evidence production; CP5 then evaluates constraint satisfaction using the resulting evidence. End-to-end test-process time is

\begin{equation}
T_{\mathrm{E2E}}=t_{\mathrm{record\ end}}-t_{\mathrm{request}}.
\end{equation}

The final system record required to determine the case outcome is formed at $t_{\mathrm{record\ end}}$. Thus, $T_{\mathrm{E2E}}$ measures the interval from the request to completion of the records needed for outcome interpretation. E2E denotes end-to-end. This interval includes the workflow and record compilation that actually occurred; subsequent researcher reading and judgment are separate. Each time metric uses applicable records containing the required timestamps. Simulation-verification time is not applicable to cases that did not enter simulation. The three metrics describe distinct observation intervals.

\subsubsection{Simulation Metrics}

Simulation evaluates candidate executability, placement outcomes, and time costs within the specified scene. Sorted instruments should be placed at designated locations; unsuitable operations may leave them away from their targets or outside the robot's reachable workspace. Placement is checked using the Euclidean distance between actual and target positions in the XY plane. For placement record $i$, the planar error is

\begin{equation}
e_i=\sqrt{(x_i-x_i^*)^2+(y_i-y_i^*)^2}.
\end{equation}

Here, $(x_i,y_i)$ is the actual instrument position and $(x_i^*,y_i^*)$ is the target position in the same coordinate system; the superscript $*$ denotes a target. Coordinates and errors are measured in meters. A record passes if $e_i$ is within the 0.06 m tolerance. The placement-validation pass rate for an execution is

\begin{equation}
R_{\mathrm{storage}}=\frac{N_{\mathrm{passed}}}{N_{\mathrm{validated}}}\times100\%.
\end{equation}

The number of records passing the placement check is $N_{\mathrm{passed}}$, and $N_{\mathrm{validated}}$ counts records that actually completed that check. Instruments without completed checks are excluded from the denominator and are not counted as passes. This metric must therefore be interpreted separately from evidence completeness. The reported mean is the arithmetic mean of the rates for the simulation evidence records, rather than a pooled rate over all instruments.

Delays associated with human intervention also affect line efficiency. Configured events represent conditions such as instrument stacking or potential entanglement that require intervention; a strategy may pause or defer the affected tasks. This design evaluates the effect of intervention time; physical testing of entanglement recognition remains future work. Intervention is represented through arrival, handling, and recovery parameters, with downtime defined as

\begin{equation}
\Delta t_{\mathrm{stop}}=(t_{\mathrm{arrive}}-t_{\mathrm{detect}})+\Delta t_{\mathrm{fix}}+\Delta t_{\mathrm{resume}}.
\end{equation}

The symbol $t$ denotes a timestamp. An anomaly is detected at $t_{\mathrm{detect}}$, and the operator arrives at $t_{\mathrm{arrive}}$; their difference is the arrival waiting time. The handling duration is $\Delta t_{\mathrm{fix}}$, and the delay before operations resume is $\Delta t_{\mathrm{resume}}$. Their sum with arrival waiting time gives $\Delta t_{\mathrm{stop}}$. All durations are measured in seconds. Default arrival, handling, and recovery delays are 1.0, 3.0, and 1.0 s, respectively. Individual cases may specify other values to test the conversion of timing requirements.

\section{Results}

An automated-script verdict is a programmatic judgment produced by applying predefined conditions to execution data. A retrospective case-review verdict is based on comparing the original input, expected conditions, actual response, stopping point, and verification evidence.

\subsection{Workflow Evaluation by Test Group}

\paragraph{TC1: Intent to Executable Strategy}

Three of the eight TC1 cases met their complete case-specific expectations (37.5\%). Some cases stopped before CP4, while others failed to meet expectations because simulation or strategy evidence was incomplete. Table~\ref{tab:9} in the Appendix summarizes the failures. Among the conforming cases, T01 completed strategy generation, simulation, constraint checking, and selection. T02 and T05 were interpreted correctly, but their strategies failed throughput constraints. Blocking those strategies was the expected outcome. These cases are counted separately from the invalid-input injections in TC4.

\paragraph{TC2: Configuration, State, and Evidence Queries}

Five of the nine TC2 query cases met expectations (55.56\%). T09, T11, T12, T13, and T14 met expectations by returning the requested placement-validation pass rate, line throughput-target settings, or task-requirement tables. T10 omitted the reset-completion value and an indication of incomplete data. T15 and T17 misclassified queries as task changes, while T16 failed to restrict its answer to the requested line.

\paragraph{TC3: Simulation, Throughput, and Timing Evidence}

Both TC3 cases met expectations, based on simulation outcomes, throughput indicators, and evidence completeness. T18 and T19 completed the virtual production-line scenarios, retained strict placement validation, and met the throughput target. Both runs produced the required throughput and placement evidence.

\paragraph{TC4: Invalid-Input Interception}

Seven of the eight TC4 invalid inputs were correctly intercepted (87.5\%). T20 and T21 prompted for a missing operator identifier and change reason, respectively. T22 routed malformed text to a help response without entering approval, simulation, or deployment. T23, T25, T26, and T27 intercepted an unsupported instrument, contradictory line scope, negative throughput target, and invalid intervention mode, respectively. Although T22 and T27 were intercepted at stages different from those anticipated, they met the objective of preventing invalid input from entering execution. T24 was the only nonconforming case: it failed to intercept an invalid line identifier.

\paragraph{TC5: Lifecycle and Outcome Records}

TC5 passed its single case, which assessed whether lifecycle events could be traced and timing metrics reconstructed. In T28, preserved n8n execution data, a strategy batch, and Isaac logs yielded a waiting time of 13.775 s, simulation-verification time excluding startup of 130.693385 s, and closed-loop time of 597.682 s. The result supports the traceability of timing evidence for this case.

\paragraph{TC6 and TC7: Model Generation}

TC6 and TC7 completed the planned generation tests. Structural validity, output consistency, and semantic correctness are reported separately rather than reduced to a single case-level pass or fail. In TC6, all three Gemma outputs were valid JSON, but none passed strict semantic checking. In TC7, two of the three Qwen outputs passed (66.67\%), whereas none of the Gemma or Llama outputs passed. TC6 reuses the three Gemma outputs included in TC7, giving nine distinct outputs in total.

\subsection{Metric-Based Evaluation}

Table~\ref{tab:6} reports requirement handling, strategy verification, and execution efficiency using the population appropriate to each metric. Excluding TC6 and TC7, 18 of 28 cases met expectations, including correct rejection of invalid inputs or infeasible strategies. Three of ten strategy-workflow requests completed autonomously (30\%), while three of eight batches entering multi-strategy simulation yielded an eligible strategy (37.5\%). The system therefore completed strategy generation and evidence verification for some requests, but autonomous completion of the full workflow remains limited. Five of nine queries succeeded (55.56\%), and seven of eight invalid inputs were correctly rejected (87.5\%). Invalid-input handling was more consistent within these cases, whereas query classification, answer content, and scope require improvement.

The mean placement-validation pass rate across eight simulation evidence records was 97.50\%, describing placements for which coordinate checks were completed. Mean times to the first reviewable response, simulation verification, and completion of the end-to-end test process were 12.94, 164.39, and 607.69 s, respectively. The system can provide requirement confirmations, query responses, or clarification relatively early, but complete verification evidence takes longer. The workflow is therefore better suited to adjustments that allow advance simulation and review. Proportions and timing metrics use different evaluated populations and should be interpreted separately. End-to-end time includes workflow execution and record compilation, but not subsequent reading and judgment.

\begin{table}[htbp]\centering
\begingroup
\small\setlength{\tabcolsep}{4pt}\renewcommand{\arraystretch}{1.13}
\caption{Evaluation metrics, populations, and quantitative results. Proportions use their stated observation units; checkpoint results are detailed in Table~\ref{tab:8}.}\label{tab:6}\medskip
\begin{tabular}{@{}>{\raggedright\arraybackslash}p{0.24500\dimexpr\linewidth-4\tabcolsep\relax}>{\raggedright\arraybackslash}p{0.59500\dimexpr\linewidth-4\tabcolsep\relax}>{\raggedright\arraybackslash}p{0.16000\dimexpr\linewidth-4\tabcolsep\relax}@{}}
\toprule
\textbf{Metric} & \textbf{Population and criterion} & \textbf{Result} \\
\midrule
Checkpoint pass rate & CP0-CP5: cases actually evaluated; CP6: four human-review records, with 26 records not reaching that stage & Table~\ref{tab:8} \\ \addlinespace[3pt]
Case-conformance rate & All 28 workflow cases, excluding TC6 and TC7; expected output and stopping behavior assessed retrospectively; correct rejection may conform & 64.29\% (18/28) \\ \addlinespace[3pt]
Autonomous strategy-workflow success rate & Ten task-change or strategy-workflow requests; applicable steps completed without manual correction; retrospective interpretation assessed separately & 30\% (3/10) \\ \addlinespace[3pt]
Candidate-batch selection rate & Eight batches entering multi-strategy simulation; at least one eligible strategy selected using CP4 evidence and CP5 constraints & 37.5\% (3/8) \\ \addlinespace[3pt]
Mean placement-validation pass rate & Arithmetic mean of eight evidence-record rates; each uses completed instrument coordinate checks with an XY tolerance of 0.06 m, assessed at CP5 & 97.50\% \\ \addlinespace[3pt]
State and configuration query success rate & Nine queries; CP1 classification and CP3 semantics assess routing, response content, and scope & 55.56\% (5/9) \\ \addlinespace[3pt]
Correct rejection rate & Eight negative inputs; correct interception assessed against the specified reason and stopping point & 87.5\% (7/8) \\ \addlinespace[3pt]
Mean time to first reviewable response & Applicable records with request and first interpretable-response timestamps; responses may confirm, answer, reject, or clarify & 12.94 s \\ \addlinespace[3pt]
Mean simulation-verification time & Applicable records with scenario-start, evidence-completion, and startup timings; CP4 execution and evidence production, excluding initial startup & 164.39 s \\ \addlinespace[3pt]
Mean end-to-end test-process time & Applicable cases with initial-request and final-required-record timestamps; workflow and record compilation included, retrospective reading and judgment excluded & 607.69 s \\ \addlinespace[3pt]
\bottomrule\end{tabular}\endgroup
\end{table}

\subsection{Evidence and Outcomes of Representative Cases}

Table~\ref{tab:7} uses six cases to illustrate how staged verification produces eligible recommendations, rejects nonconforming strategies, and reveals unresolved problems. T01 demonstrates an evidence-supported recommendation. T02 and T05 show that correct requirement conversion and successful execution do not suffice: throughput constraints must also be met before a candidate is eligible for adoption. These cases support the screening role of staged verification. T04, T06, and T24 expose gaps in evidence completeness, semantic preservation, and invalid-input interception, respectively. Case judgments must therefore consider the original request, execution evidence, and stopping reason together.

\begin{table}[htbp]\centering
\begingroup
\small\setlength{\tabcolsep}{4pt}\renewcommand{\arraystretch}{1.13}
\caption{Verification evidence and outcomes of representative cases.}\label{tab:7}\medskip
\begin{tabular}{@{}>{\raggedright\arraybackslash}p{0.23500\dimexpr\linewidth-4\tabcolsep\relax}>{\raggedright\arraybackslash}p{0.55500\dimexpr\linewidth-4\tabcolsep\relax}>{\raggedright\arraybackslash}p{0.21000\dimexpr\linewidth-4\tabcolsep\relax}@{}}
\toprule
\textbf{Case and request} & \textbf{Key evidence and subsequent handling} & \textbf{Case verdict} \\
\midrule
T01: line 1 at least 90 instruments/h & Simulation and constraint checks passed; line-1 throughput was 91.5914 instruments/h. A recommendation was provided, and deployment was declined as required by the test. & Pass: verification and review completed. \\ \addlinespace[3pt]
T02: all lines at least 120 instruments/h & All three candidates failed mandatory constraints. No strategy was selected, and the system requested revised requirements. & Pass: ineligible strategies correctly blocked. \\ \addlinespace[3pt]
T05: knife handles as the line-2 target & Target conversion was correct, but candidates failed the existing throughput constraint and were rejected. & Pass: existing constraints preserved and enforced. \\ \addlinespace[3pt]
T04: adjust arrival, handling, and recovery times & Timing values were converted correctly, but candidate execution evidence was incomplete. & Fail: evidence workflow incomplete. \\ \addlinespace[3pt]
T06: prioritize non-scissors instruments & The priority became FCFS, so subsequent simulation evaluated the wrong strategy. & Fail: requirement semantics distorted. \\ \addlinespace[3pt]
T24: set a throughput target for line -2 & The invalid line was not intercepted; the workflow stopped only when the test script cancelled it. & Fail: invalid input not correctly intercepted. \\ \addlinespace[3pt]
\bottomrule\end{tabular}\endgroup
\end{table}

\subsection{Checkpoint Results and Limitations}

Table~\ref{tab:8} presents the additional evidence provided by semantic, simulation, and constraint checks. All 12 outputs evaluated at CP2 met structural requirements, yet 13 of the 27 cases evaluated at CP3 failed semantic requirements. Requirement conversion and query scope therefore remain major sources of error. At CP4, two of nine cases failed simulation-execution checks; at CP5, three of seven cases with outcome evidence failed constraint checks. Later DT-related checks thus reveal execution and operational problems that structural checks cannot identify. For example, T04 lacked sufficient strategy evidence for a conclusive outcome, and T05 was blocked because it failed the existing throughput constraint. The workflow requires both adequate evidence and constraint satisfaction before subsequent review.

A recorded checkpoint failure does not necessarily indicate successful interception during execution. In T06, the requested priority was not preserved, yet simulation still evaluated the incorrect strategy. This is a defect revealed by evaluation, not a successful interception. Among the eight negative inputs in Table~\ref{tab:6}, seven were handled as expected, while T24 failed to intercept an invalid production line, showing a remaining gap.

All four records that completed the preceding checks, produced complete evidence, and actually reached CP6 passed human review. Thus, every result submitted for final human confirmation in this test set met the specified review requirements. Together with the correct blocking of throughput-ineligible strategies in T02 and T05, this outcome supports both the screening of ineligible strategies and the confirmation of reviewed results through staged verification. The evaluated cases demonstrate these screening and confirmation effects; overall erroneous decisions and long-term performance remain to be evaluated.

\begin{table}[htbp]\centering
\begingroup
\small\setlength{\tabcolsep}{4pt}\renewcommand{\arraystretch}{1.13}
\caption{Evaluated cases and pass rates at each checkpoint. CP0-CP5 rates divide passes by entries. CP0 reports input-validity flags; invalid-input rejection outcomes are assessed separately in TC4. CP6 uses the four records actually subjected to final human review; the remaining 26 records did not enter that stage.}\label{tab:8}\medskip
\begin{tabular}{@{}>{\raggedright\arraybackslash}p{0.07000\dimexpr\linewidth-10\tabcolsep\relax}>{\raggedright\arraybackslash}p{0.43000\dimexpr\linewidth-10\tabcolsep\relax}>{\raggedright\arraybackslash}p{0.11000\dimexpr\linewidth-10\tabcolsep\relax}>{\raggedright\arraybackslash}p{0.11000\dimexpr\linewidth-10\tabcolsep\relax}>{\raggedright\arraybackslash}p{0.11000\dimexpr\linewidth-10\tabcolsep\relax}>{\raggedright\arraybackslash}p{0.17000\dimexpr\linewidth-10\tabcolsep\relax}@{}}
\toprule
\textbf{CP} & \textbf{Evaluation focus} & \textbf{Entered} & \textbf{Passed} & \textbf{Failed} & \textbf{Pass rate (\%)} \\
\midrule
CP0 & Input validity & 30 & 22 & 8 & 73.33 \\ \addlinespace[3pt]
CP1 & Intent and required fields & 30 & 26 & 4 & 86.67 \\ \addlinespace[3pt]
CP2 & Structural validity & 12 & 12 & 0 & 100.00 \\ \addlinespace[3pt]
CP3 & Task and scenario semantics & 27 & 14 & 13 & 51.85 \\ \addlinespace[3pt]
CP4 & Simulation executability & 9 & 7 & 2 & 77.78 \\ \addlinespace[3pt]
CP5 & Outcome and constraint conformance & 7 & 4 & 3 & 57.14 \\ \addlinespace[3pt]
CP6 & Final human review & 4 & 4 & 0 & 100.00 \\ \addlinespace[3pt]
\bottomrule\end{tabular}\endgroup
\end{table}

\subsection{Exploratory Comparison of Model Generation}

Gemma, Qwen, and Llama each generated three outputs for a shared input combining a line-1 throughput target with a four-line simulation configuration. All nine outputs met structural requirements. Two of the three Qwen outputs passed strict semantic checking (66.67\%); none of the Gemma or Llama outputs passed. Gemma expanded the single-line throughput target to all four lines in every output. Qwen correctly separated the scopes twice. Llama requested additional clarification in all three outputs and did not produce the specification ready for review expected by this case.

Seven of the nine structurally valid outputs failed strict semantic checking, supporting scope and meaning checks before simulation. This comparison covers one input and three outputs per model and identifies scope confusion and unnecessary clarification. Broader requirement sets are needed to assess model performance across tasks.

\section{Discussion and Limitations}

\subsection{Anomaly Types and Observed Handling}

Staged verification handles anomalies according to their sources. Missing required information, contradictions, and some invalid values in operator instructions can be addressed before simulation. Seven of the eight negative inputs in TC4 were correctly intercepted (87.5\%). Including T08, an unsupported-instrument case from another test group, gives seven correct interceptions among nine input-anomaly cases (77.78\%). T23 explicitly identified the unsupported term, whereas T08 asked only for the intended lines, showing variation with expression and context. T24's invalid line identifier passed candidate validation, identifying a gap in line-identifier checking.

For correctly translated requirements whose candidates failed operational constraints, both T02 and T05 were rejected because of insufficient throughput. None of the six candidates across these two batches was selected. This supports the screening role of simulation and constraint checking within the generated candidate sets. For simulation or evidence-workflow anomalies, both T03 and T04 paused recommendations and requested repair because evidence was incomplete, preserving an unresolved status. The intended verification remained incomplete in both cases, and the records contain no successful post-recovery result. Pausing recommendations and completing recovery therefore require separate evaluation.

Input interception, rejection for insufficient throughput, and pausing for incomplete evidence have distinct observation units and success criteria. Future evaluation should report correct interceptions, missed interceptions, false rejections, and completed recoveries by anomaly source to characterize both protection and task completion.

\subsection{Human-AI Responsibilities and Verification Cost}

The system automates candidate generation, simulation, constraint checks, and evidence summarization, while the operator retains responsibility for requirement confirmation and strategy adoption. In this study, the researchers acted as operators and subsequently interpreted the records. Three of ten strategy-workflow requests completed autonomously, and five of nine queries succeeded, indicating continuing needs for clarification, correction, and failure diagnosis. Measuring the human effort involved in these activities would quantify the effect of evidence organization on operator workload.

Across eight distinct primary execution evidence records, mean simulation-verification times excluding startup were 127.71 s for T01, T18, and T19, which produced recommendations; 124.05 s for T02 and T05, which were rejected for insufficient throughput; 277.24 s for T03 and T04, which had incomplete evidence; and 129.38 s for T06, which entered simulation after semantic mistranslation. Success, rejection, and incomplete verification all incur computational costs. These observations motivate semantic checks before simulation and earlier identification of evidence gaps during execution. Scene configurations and group sizes differ, so the times describe the existing cases. The values for T03 and T04 refer to the cited execution evidence; verification of their complete candidate batches remained unfinished.

Mean times to the first reviewable response and completion of the end-to-end test process were 12.94 and 607.69 s, respectively. The first response may request clarification, while end-to-end time includes workflow execution and record compilation; retrospective researcher interpretation was not timed. The workflow is suited to adjustments that allow advance simulation. Matched tasks should be used to compare anomaly recognition, stopping, and recovery times, as well as total completion time, human effort, and judgment accuracy against manual configuration.

\subsection{Semantic Consistency and Bounded Candidate Sets}

Consistency between original intent and execution specifications is a primary improvement area. T06 translated a non-scissors-first request into first-come, first-served. Although simulation reported zero priority-rule deviation, the executed specification had already diverged from the intended rule. The case was ultimately rejected for insufficient throughput, demonstrating enforcement of that constraint while the semantic error passed earlier checks. Simulation can execute an incorrect specification consistently. Line scope, target instruments, and priority rules should therefore be checked directly across the original request, structured representation, and scenario specification.

The direct model comparison supports this observation: all nine outputs met structural requirements, but only two passed strict semantic checking. The remaining outputs comprised four scope expansions and three unnecessary clarification requests. This comparison covers one input and three generations per model, identifying failure modes under a specific requirement. Query-routing errors and omitted response content in workflow cases may involve the model, prompts, or data interfaces; preserving each module's inputs and outputs would help locate their causes.

Strategy selection is bounded by three candidates per batch and the specified scene constraints. A selected strategy is eligible and better ranked within that batch. Future work should expand requirements and candidate sets and compare generation and ranking methods under matched conditions.

\subsection{Within-Simulation Deviations and Digital-Twin Fidelity}

The CSSD scene evaluates instrument sorting, placement, and parameterized intervention. The 97.5\% mean placement-validation pass rate is based on eight distinct execution evidence records, with an XY position tolerance of 0.06 m, and measures placements whose checks were completed within simulation. Eight of ten placement records in T03 passed (80\%), showing that the system measures within-simulation deviations. Its final stopping reason was incomplete candidate evidence; the independent interception effect of placement deviations requires separate testing.

Digital-twin fidelity concerns agreement between simulation predictions and physical behavior. Physical comparisons and tests with known model biases are still needed to measure cases judged feasible in simulation that fail physically. If later checks share biased simulation results, multiple checks may accept the same erroneous evidence. Independent physical references or observations should calibrate positions, contact behavior, and intervention times before measuring how model bias affects strategy decisions. Contact forces, instrument damage, sterilization quality, and physical entanglement recognition also require their own validation evidence.

\subsection{Evaluation Credibility and Reliability Improvement}

The evaluation judgments themselves require validation. Among the 28 workflow cases, four of the eighteen script passes failed retrospective outcome interpretation, while three correct responses were marked as failures by the script. Comparing original requests, actual responses, and stopping reasons helped identify these discrepancies. Independent annotation would allow assessment of judgment consistency. The four CP6 passes were T01, T18, T19, and T28; T28 shares the execution evidence and candidate batch of T01, so these four test records correspond to three distinct executions.

The correct interception rate of the tested invalid inputs was 87.5\%, demonstrating an input-screening effect; the framework's effect on overall erroneous decisions remains to be evaluated. Eighteen of the 28 workflow cases met expectations (64.29\%), characterizing the complete framework on the fixed test set. Future matched comparisons can hold requirements and candidate sets constant while adding or removing input validation, original-intent checking, simulation verification, and human review. Correct interception, erroneous release, false rejection of valid requests, and completed recovery should be measured by anomaly type, together with time and human effort. Because checkpoints share requirement representations or simulation data, their contributions should be estimated with these dependencies in view. Expanded evaluation should specify included records, repeated executions, and missing-data rules to assess long-term performance and applicability across settings.

\section{Conclusion}

This study developed an intent-review framework combining proposals from an LLM, verification in a digital twin, and human decision-making, implemented for multi-line task adjustment. Operator intent, requirement versions, candidate strategies, scenario specifications, and execution evidence are linked. Mandatory constraints filter a bounded candidate set before a recommendation is issued, providing traceable grounds for both adoption and rejection.

Of the 28 workflow cases in the virtual CSSD setting, eighteen met expectations (64.29\%); three of ten strategy-workflow requests completed autonomously (30\%); and seven of eight invalid inputs were correctly intercepted (87.5\%). All four records that completed preceding checks, produced complete evidence, and reached CP6 passed human review. Together with the rejection of throughput-ineligible strategies, these results demonstrate the screening and confirmation effects of staged review within the fixed test set. Semantic distortion, evidence gaps, and missed invalid lines are priorities for improvement; overall erroneous decisions and long-term performance remain to be evaluated.

The study provides an implemented, traceable review framework and fixed-case evaluation. Physical-deployment outcomes and human effort remain to be measured. Broader requirement sets, matched comparisons, operator studies, and physical evaluations will assess effects on decision quality, human effort, and production performance.

\clearpage\appendix
\section*{Appendix}

\begin{table}[htbp]\centering
\begingroup
\footnotesize\setlength{\tabcolsep}{4pt}\renewcommand{\arraystretch}{1.13}
\caption{Main failure cases and retrospective review findings.}\label{tab:9}\medskip
\begin{tabular}{@{}>{\raggedright\arraybackslash}p{0.05500\dimexpr\linewidth-10\tabcolsep\relax}>{\raggedright\arraybackslash}p{0.11000\dimexpr\linewidth-10\tabcolsep\relax}>{\raggedright\arraybackslash}p{0.14500\dimexpr\linewidth-10\tabcolsep\relax}>{\raggedright\arraybackslash}p{0.12500\dimexpr\linewidth-10\tabcolsep\relax}>{\raggedright\arraybackslash}p{0.25000\dimexpr\linewidth-10\tabcolsep\relax}>{\raggedright\arraybackslash}p{0.31500\dimexpr\linewidth-10\tabcolsep\relax}@{}}
\toprule
\textbf{Case} & \textbf{Failure stage} & \textbf{Failure source} & \textbf{Script verdict} & \textbf{Rejection reason or observed handling} & \textbf{Retrospective review finding} \\
\midrule
T03 & Simulation & DT scenario issue & Pass & Candidates lacked conclusive simulation evidence. & Two lines, immediate stop, and five instruments per line were represented correctly, but DT verification was not completed. \\ \addlinespace[3pt]
T04 & Evidence pipeline & Simulator or API error & Pass & Complete strategy evidence was not produced. & Timing fields were converted correctly, but strategy evidence remained incomplete. \\ \addlinespace[3pt]
T06 & Semantic check & Priority rule not enforced & Fail & Semantic mistranslation passed earlier checks; throughput failure caused later rejection. & The non-scissors-first rule for line 1 became first-come, first-served (FCFS), so simulation evaluated the wrong strategy. \\ \addlinespace[3pt]
T07 & Intent routing & Query confused with task change & Fail & Incorrectly entered candidate approval. & A request for a 99-line requirement table was treated as a production change, without obtaining line definitions or returning the table. \\ \addlinespace[3pt]
T08 & Intent check & Unsupported instrument not identified & Pass & Asked only which lines were intended. & Clarification did not identify "unicorn clamps" as unsupported, failing the expected clarification requirement. \\ \addlinespace[3pt]
T10 & Query response & Evidence value omitted & Fail & No rejection recorded; required value absent. & The reply described the latest run but omitted reset completion and an incomplete-data notice. \\ \addlinespace[3pt]
T15 & Intent classification & Query confused with task change & Inconclusive & Requested task-change fields. & A line-3 state query became a task-change request; no structured tool trace was captured for script scoring. \\ \addlinespace[3pt]
T16 & Query response & Line filter omitted & Fail & No rejection recorded; wrong response scope. & A line-2 state query received global arrival-time settings rather than the requested line record. \\ \addlinespace[3pt]
T17 & Intent classification & Query confused with task change & Inconclusive & Routed to task-change handling. & The reset-completion query became a task change rather than a report query; no structured tool trace was captured. \\ \addlinespace[3pt]
T24 & Intent check & Invalid line not intercepted & Pass & No correct rejection; the test script cancelled. & Line -2 passed candidate validation. Cancellation by the script does not count as system interception. \\ \addlinespace[3pt]
\bottomrule\end{tabular}\endgroup
\end{table}

\section*{Funding}

This work was supported by the National Science and Technology Council, Taiwan, under Grant NSTC 115-2218-E-008-008, and by an industry-academia collaboration project with Compal Electronics, Inc. The funders provided financial support only and had no role in study design, data collection, analysis or interpretation, manuscript preparation, or the decision to submit the article for publication.

\section*{Declaration of Competing Interests}

The authors declare that they have no known competing financial interests or personal relationships that could have appeared to influence the work reported in this paper.

\section*{Data and Code Availability}

The data and code supporting this study are available from the corresponding author upon request.

\section*{CRediT Author Statement}

Teng-Hsien Ko: Software, Investigation, Data curation, Formal analysis, Writing - original draft. Chin-Te LIN: Conceptualization, Funding acquisition, Writing - review and editing.

\section*{Declaration of Generative AI and AI-Assisted Technologies in Manuscript Preparation}

The authors developed the research questions, designed the method, implemented the system, conducted the experiments, acquired the data, and performed the initial analyses. During manuscript preparation, the authors used OpenAI Codex to assist with organization, presentation of existing research records, translation, and language editing. The authors reviewed and revised the manuscript and take full responsibility for its content.

\end{document}